**Free-space optomechanical liquid probes using a twin-microbottle resonator**

Motoki Asano,[1]* Hiroshi Yamaguchi,[1] and Hajime Okamoto[1]
1. NTT Basic Research Laboratories
*motoki.asano@ntt.com

**Abstract**

Cavity optomechanics provides high-performance sensor technology, and the scheme is also applicable to liquid samples for biological and rheological applications. However, previously reported methods using fluidic capillary channels and liquid droplets are based on fixed-by-design structures and therefore do not allow an active free-space approach to the samples. Here, we demonstrate an alternate technique using a probe-based architecture with a twin-microbottle resonator. The probe consists of two microbottle optomechanical resonators, where one bottle (for detection) is immersed in liquid and the other bottle (for readout) is placed in air, which retains excellent detection performance through the high optical-$Q$ (~$10^7$) of the readout bottle. The scheme allows the detection of thermomechanical motion of the detection bottle as well as its optomechanical sideband drive. This technique could lead to *in-situ* metrology at the target location in arbitrary media, and could be extended to ultrasensitive biochips and rheometers.

**MAIN TEXT**
**Introduction**
The light confined in an optical microcavity affects the vibration of the cavity and vice versa, via radiation pressure when the cavity simultaneously forms a mechanical resonator with vibrational degrees of freedom. Such a cavity optomechanical structure can be utilized in a variety of optomechanical devices, such as sensors, actuators, filters, transducers, circulators, and amplifiers (1). Their performance is significantly improved when the optomechanical structure has a high optical and mechanical quality factor ($Q$), which depends on the surrounding conditions in addition to the inherent quality of the material. Generally, state-of-the-art optomechanical devices show high-$Q$ optical and mechanical modes in air and vacuum conditions. Their high-$Q$ conditions provide excellent optomechanical performance, such as high displacement sensitivity (2), high wavelength conversion efficiency (3), and low cooling limit (4).

The applicable condition of cavity optomechanical structures is not limited to a vacuum and gas atmosphere, but extends to liquid, which is the focus of interest for metrological applications in biology and rheology. A mechanical resonator is damped in liquid and frequency shifted due to the surrounding viscous medium. This property can be used to optomechanically monitor the viscosity and density of the medium (5-7). Moreover, if some molecules or particles adhere to a mechanical resonator, the mechanical resonance changes. Thus, the optomechanical probe also allows chemical and biological sensing in liquid (8-12). However, the sensitivity in liquid is orders of magnitude lower than that in air, because the optical $Q$ deteriorates due to the light loss to the surrounding medium having a higher refractive index than the optical cavity (11,13). Therefore, some contrivance that can suppress the optical loss is required to achieve high sensitivity for the cavity optomechanical approach in liquid.

Recently, two innovative optomechanical approaches have been demonstrated to improve the optical $Q$ in liquid (5,8,10,12). The first approach is based on a glass microcapillary. The capillary consists of a hollow-core optomechanical microbottle (5,6,8), which forms both the optical whispering gallery modes (WGMs) and mechanical radial breathing modes (RBMs). The liquid flowing inside the core influences the mechanical RBMs and affects the optical WGMs through optomechanical coupling via radiation pressure. This allows optical detection of mass flow and viscoelasticity in the fluidic channel (6). Because the optical WGMs are formed on the outer surface (in air) and isolated from the liquid flowing inside the channel, the high optical $Q$ is conserved, enabling an optical detection of the thermal motion and an optical drive of the mechanical RBMs. The second approach is based on a droplet-covered GaAs microdisk (7,10,12), which forms the optical WGMs and the mechanical RBMs. Because the refractive index of GaAs (3.3 at telecom wavelengths) is higher than that of typical viscous media, the light is confined in the optical cavity, leading to a high optical $Q$ in liquid ($>10^5$). The optomechanical detection of viscoelasticity (7) and a single bacterium (12) in liquid droplets has been demonstrated by using GaAs microdisks.

The above two approaches fit cases where the liquid portion is injectable and/or certainly includes the target specimen, however, neither approach allows versatile operations in liquid, because both are based on fixed-by-design structures. To widely extend the cavity optomechanics in liquid, an alternative scheme that allows free spatial access to liquid is desired. Such a probe-based architecture in similar to the scanning probe microscope with a mechanical cantilever (15,16) would enable *in-situ* cavity optomechanical metrology at the target location in liquid, leading to the study of local dynamics and rheology in fluid.

Here, we demonstrate cavity optomechanics using a twin-microbottle resonator (TMBR) that enables active free spatial access to liquid. The TMBR consists of two microbottles (see Fig.1A) that form both the optical WGMs (14-25) and the mechanical RBMs (26-28). The WGMs and RBMs are strongly coupled via radiation pressure. Therefore, an evanescent optical access to the WGMs with a telecom-wavelength tapered fiber allows optical drive and detection of the mechanical RBMs through the optomechanical coupling. The important characteristic in this TMBR is that the mechanical RBMs in the two bottles are elastically interconnected through the neck part, while the optical WGMs in the two bottles are isolated. This mechanically coupled but optically isolated configuration (see Fig.1B) allows high-performance optomechanical operation in liquid when one bottle (for detection) is immersed in liquid and the other bottle (for readout) is placed in air. The mechanical motion of the bottle in liquid can be read out through the optical modes of the bottle in air with a high $Q$. Thus, a highly sensitive detection of mechanical motion of the microbottle, including the monitor of thermal fluctuations, is achieved in liquid, where one can freely target the probing position and adjust the probe's immersion depth by using a micropositioner. The high optical $Q$ also allows cavity optomechanical operation in the resolved sideband regime, where the frequency of the mechanical RBMs exceeds the linewidth of the optical WGMs. The pump light blue-detuned from the optical cavity resonance causes optomechanically induced amplification (29,30) of the damped mechanical motion in liquid, leading to enhanced displacement and enabling the phase information to be utilized for phase-locked-loop detection.

## Results
### Fundamental properties of TMBR in air
The TMBR is fabricated from a silica optical fiber with the clad diameter of 125 µm by using the heat and pull technique. A fine adjustment of the fiber tension while changing the heating position allows us to form three neck parts of 115 µm in diameter (indicated by white arrows in Fig. 2B), which separate two 720-µm-long microbottles (the upper and lower bottles) with the maximal diameter of 125 µm (indicated by red arrows in Fig. 2B) . Because the difference in diameter between the bottle and the neck (~10 µm) is larger than the optical wavelength (1.55 µm), the optical WGMs in the two microbottles are isolated, while the mechanical RBMs are coupled with each other (see Supplementary Material). To optically access to the WGMs, a tapered fiber with a diameter close to the optical wavelength is orthogonally contacted to one of the microbottles (Fig. 2A). This allows an evanescent optical coupling to the WGMs, which is confirmed by the optical transmission spectrum (see Methods for details of experimental setup). In air, some of the WGMs have the optical $Q$ on the order of $10^7$, which is confirmed via the linewidth of the transmission resonances (Fig. 2C). The high-$Q$ optical mode used in this study ($= 1.65 \times 10^7$), which corresponds to the typical value in the reported high-$Q$ microbottle resonators (17-28), enables cavity optomechanical operation in the resolved sideband regime, i.e., the optical linewidth ($= 12.1$ MHz) is smaller than the mechanical mode frequencies ($=31$ MHz). The coupled mechanical RBMs in the TMBR can be probed via the high-$Q$ optical WGMs in one of the microbottles (hereafter in the upper microbottle placed in air) through optomechanical coupling via radiation pressure (see Fig. 1B).

First, we characterized the mechanical resonances of the TMBR in atmospheric conditions (in air). In the twin-microbottle structure, several sets of spilt RBMs appear around 31 MHz, which correspond to the coupled RBMs with no node along the radial direction but with different node numbers with respect to the axial direction (see Methods and Supplementary Material). Of these sets, we focus on the one that provides maximal optomechanical transduction efficiency, i.e., the highest signal level. Figure 2D shows the thermal noise spectrum of the coupled mechanical RBMs measured with the probe light of 1.8 mW. This is a measure of the frequency noise of the optical transmission when the detuning of the probe light is set on the slope of the cavity resonance, where the optical transmission periodically modulates at the mechanical mode frequency via the optomechanical coupling. The high-frequency mode with the higher signal level ($\Omega_h/2\pi = 31.41$ MHz) corresponds to the anti-symmetry RBM, in which the upper bottle breathes in the opposite direction to the lower bottle (Fig. 2E). Note that the amplitude of the upper bottle, which directly couples to the optical WGM, is larger than that of the lower bottle because of the intrinsic eigenfrequency difference between the two bottles ($\Omega_L < \Omega_U$). On the other hand, the low-frequency mode with the lower signal level ($\Omega_l/2\pi = 31.31$ MHz) corresponds to the symmetry RBM, in which the upper and lower bottles breathe (Fig. 2E) in the same direction, similar to coupled micromechanical resonators (31,32). In this mode, the amplitude of the lower bottle is larger than that of the upper bottle. The mechanical damping rates, which corresponds to the full-width-at-half-maximum of the mechanical resonance, are $\Gamma_h/2\pi = \Gamma_l/2\pi = 11.1$ kHz for both the high-frequency and low-frequency modes, leading to the mechanical quality factor of $Q_h = Q_l = 2.8 \times 10^3$. The intrinsic eigenfrequency difference ($\delta_M = (\Omega_U - \Omega_L)/2\pi$) between the two microbottles and the mechanical coupling strength ($g_M$) are extracted by fitting the two resonances with the coupled resonator model (see Methods), where $\delta_M = 75.1$ kHz and $g_M/2\pi = 59.6$ kHz. The coupling strength exceeds the damping rate of the each mode, i.e., $g_M/\Gamma_h = g_M/\Gamma_l = 5.3 > 1$, leading to the strong mechanical coupling. The strong-coupling and low-dissipation nature provide efficient transduction between the two mechanical modes with mechanical cooperativity exceeding the unity ($C_M = 9.3 > 1$).

The vacuum optomechanical coupling rate is extracted by using a phase-modulated probe signal from an electro-optic modulator (EOM) (33). The resulting value is $g_{0,h}/2\pi = 1.3$ kHz for the high-frequency mode and $g_{0,l}/2\pi = 0.7$ kHz for the low-frequency mode. The signal-to-noise ratios (i.e., the ratios between the peak top level and the noise floor level) are 3.2 and 1.7, which result in displacement sensitivity (i.e., the minimum detectable displacement) of $4.5 \times 10^{-18}$ m/$\sqrt{\text{Hz}}$ in our optomechanical setup (see Methods). This sensitivity is on a level similar to that in microdisks (7,10,12) and microcapillaries (5,6,8), and will be further improved by constructing a balanced homodyne interferometer (33). Combining the efficient transduction between two mechanical modes and the strong optomechanical coupling enables us to detect the thermal motion when the TMBR is immersed in liquid.

**Optical probe of the thermal motion of the microbottle in water**

This TMBR can be dipped into liquid at the target location while changing the immersion depth with a micropositioner (Fig. 3A). When the lower microbottle is partially immersed in water, the eigenfrequency $\Omega_L$ shifts to the lower frequency side because the effective mass of the mechanical mode increases in the condition where the lower microbottle is surrounded by the water molecules (5). This leads to a red shift of the low-frequency mode ($\Omega_l$), as shown in Fig. 3B. Figure 3C shows the experimental result measured with the probe power of 1.8 mW when the lower microbottle is partially immersed in water in the beaker by changing the depth step by step. Here, the vertical axis corresponds to the relative immersion depth $z_{im}$, which is defined as the depth from the initial position of the water surface close to the bottom neck (indicated by the bottom white arrows in Fig. 2B). In the very early stage of immersion (i.e., when $z_{im} < 100$ μm), the coupled mechanical modes do not show the apparent change because the water surface is still below the edge of the mechanical mode distribution. At $z_{im} \cong 140$ μm, a local peak appears in the frequency shift ($\delta\Omega$) and the linewidth change ($\delta\Gamma$) (Fig. 3E and 3F). This is found only in the low-frequency mode and is evidence that the mechanical mode and water start to overlap. When $z_{im} > 300$ μm (i.e., at least a third of the lower microbottle is in water), a significant change appears in the low-frequency mode. $\Omega_l$ decreases with increasing $z_{im}$ (Fig. 3E). The linewidth, which corresponds to the damping rate ($\Gamma_l/2\pi$), also increases with increasing $z_{im}$ in this regime (Fig. 3F). This is due to the viscous damping in liquid, which was previously reported in liquid optomechanics (7). At $z_{im} = 500$ μm, $\Gamma_l/2\pi$ reaches 56.1 kHz, resulting in degradation of the mechanical quality factor of $Q_l = 5.6 \times 10^2$. Note that the high-frequency mode does not show a large change even in this condition, because it is dominated by the upper microbottle in air. Only a slight shift appears in $\Omega_h$ at $z_{im} \approx 500$ μm, which is a consequence of the large shift in $\Omega_L$ (Fig. 3E). Because the upper microbottle is still in air and therefore maintains the high optical $Q$ ($= 1.65 \times 10^7$), the displacement sensitivity for the low-frequency mode at $z_{im} = 500$ μm stays at the same level ($\sim 5 \times 10^{-18}$ m/$\sqrt{Hz}$) as in air for the same probe power (1.8 mW), where the sensitivity (i.e., the noise floor level) depends on the probe power because of the power-dependent optomechanical coupling and transduction efficiency (Fig. 3D).

The experimentally measured values of the frequency shift and the linewidth broadening are fit well with the theoretical model assuming an incompressible medium, which was previously applied to droplet-covered GaAs (7), with the help of the analytical expression of the mechanical mode distributions in the microbottles (see Methods). The good agreement between the experiments and theory, except for the local peak at $z_{im} \cong 140$ μm (see broken curves in Fig. 3E and 3F), allows the extraction of the viscous damping rate and eigenfrequency shift in the lower microbottle in water (see Methods). A more detailed theoretical model, one that includes the liquid-air interface effect, will be necessary to reproduce the overall regime for the immersion depth including the local peak at $z_{im} \cong 140$ μm. Such an advanced theoretical approach in combination with the optomechanical measurements using this TMBR will open up intriguing research on the phenomena at the liquid-air interface.

**Cavity optomechanical operation in resolved sideband regime**
To improve the signal visibility, we performed a driven measurement of the mechanical modes. In addition to the displacement amplitude, the external driving provides phase information, which allows temporal tracking of the mechanical resonance with a phase-locked loop (34,35). The present TMBR enables such optical driving of the mechanical motion in liquid using the framework of *cavity optomechanics* (1) in the resolved sideband regime ($\kappa_U/\Omega_h$ ($\kappa_U/\Omega_l$) = 2.6 (2.5) > 1. This is demonstrated by introducing the frequency-detuned pump light in addition to the probe light. Here, the pump light is injected by modulating the frequency of the laser light used for the probe light with EOM, as shown in Fig. 4A. The transmission $S_{21}$ signal of the probe light is detected by a vector signal analyzer. When the pump light is red-detuned from the cavity on-resonance by $\Delta \approx -\Omega_l$, it results in optomechanically induced transparency (OMIT), leading to the hybridization of the mechanical mode and optical mode (29,30,36). On the other hand, when the pump light is blue-detuned from the cavity on-resonance by $\Delta \approx +\Omega_l$, it results in optomechanically induced amplification (OMIA), leading to the excitation of the mechanical motion (29,30) (Fig. 4B). OMIA is especially useful in enhancing the mechanical response and obtaining the phase information.

Figure 4C shows the frequency response of the two mechanical modes measured through the transmission ($S_{21}$) signal of the probe light for three different blue-detunings at $z_{im} = 0$ μm, where the detuning value Δ was extracted from the theoretical fitting (broken curves). When $\Delta \leq 0.7\Omega_{h,l}$ (red and green plots), the pumping frequency is further detuned from the sum of the mechanical and optical mode frequencies; thus, the optical driving of the mechanical mode causes the Fano-type weak response. In contrast, when the detuning is close to the mechanical mode frequencies, i.e., $\Delta = 0.9\Omega_{h,l}$ (the blue plot), the mechanical modes are optically excited to show the Lorentzian-type response, leading to the sufficient excitation of the mechanical modes. This optomechanical driving can be utilized even when the lower microbottle is partially immersed in water. Figure 5A shows the $z_{im}$ dependence of the frequency response of the two mechanical modes when $\Delta = 0.9\Omega_{h,l}$. Here, one can clearly see the change in the resonance frequency and the linewidth for the low-frequency mechanical mode with respect to the immersion depth, where the amount of the change is similar to that in the case of the thermal motion (in Fig. 3C). The phase information additionally obtained by this driven measurement allows the construction of a phase-locked loop, which is practically useful for tracing the change in the mechanical resonance in time induced, for example, by particle absorption and chemical reactions, which is necessary for the *in-situ* metrological applications in liquid (35).

**Discussion**
The unique characteristic of this new optomechanical architecture is its ability to freely access the target location in liquid in similar to scanning probe microscopy (15,16). This ability will open up new dimensions of cavity optomechanics in liquid, such as the sensing of non-uniformly distributed molecules and particles in liquid, detection of the liquid-liquid interfaces in mixed liquids with difference viscosities, and monitoring of turbulent liquid flow throughout the detection of the liquid-air interface. Sensing in highly viscous media and non-Newtonian liquids, e.g., including macromolecules and supramolecules, could also be available with the help of optomechanically induced amplification of the heavily damped mechanical modes using the high-*Q* optical cavity.

This glass-based architecture allows partial metal coating and surface chemical modification. This will enable selective detection of biochemical species and biomolecules through molecular recognition, similar to how the previously reported optical WGM biosensors do (13). The coating material can be chosen freely regardless of its optical refractive index and absorbance because of the optical isolation between the two microbottles. Furthermore, owing to its scalable glass fabrication technique, multiple microbottles can be interconnected to form arrayed structures. By using the multiple microbottle structures in combination with selective surface modification, it will be possible to realize highly sensitive biochips, such as DNA and protein chips (37-39).

In summary, cavity optomechanics in water using a twin-microbottle resonator has been demonstrated. The optical high-*Q* modes in the microbottle placed in air enables highly sensitive detection of thermal fluctuation and resolved-sideband operation of the microbottle partially immersed in water. The applications of this new optomechanical architecture are not limited to the condition in water but can be widely extended to the condition in arbitrary liquids, even in highly viscous ones and gel-like media. The free accessibility to the target location would lead to pioneering research on phenomena in non-uniform media and/or at liquid-air and liquid-liquid interfaces using the framework of cavity optomechanics.

**Materials and Methods**
**Experimental setup for tracking thermal fluctuations**
A schematic of the measurement setup is shown in Fig. 2A. An external diode laser (ECDL) with an optical wavelength of 1550 nm was used to probe thermal fluctuations in the twin microbottle resonators. The probe light from the ECDL was sent to an erbium-doped fiber amplifier (EDFA), an electro-optic modulator (EOM), a polarization controller, and a variable optical attenuator (VOA). After appropriately adjusting the optical power at 1.8 mW and the polarization, the probe light propagates on a tapered optical fiber with the dimeter of about the wavelength, which enables us to efficiently couple to an optical WGM on the microbottle in air. The output light from the twin microbottle resonators, which contains the phase-modulated signals owing to the optomechanical interaction, was detected by an avalanche photodiode (APD). The DC component of electric signal from the APD was monitored by a digital storage oscilloscope (DSO), and the AC component was detected by an electric spectrum analyzer (ESA). The transmission spectrum (shown in Fig. 2C) was observed by scanning the laser frequency in the ECDL and monitoring the transmission in the DSO. The power spectral density of thermal fluctuations in the coupled mechanical mode (shown in Fig. 2D) was measured in the ESA by fixing the laser frequency via the thermal locking method. In the same manner, the power spectral density of thermal fluctuations with respect to the immersion depth of water was measured in the ESA with the thermal locking method. The immersion depth was controlled by a stepping motor with a single step of 7.8 μm. To quantify the optomechanical coupling, a calibrating electro-optic tone at the frequency of 31.5 MHz was generated in the EOM.

**Analytical expression of mechanical vibrations in microbottle resonator**
The analytical expression of spatial distribution of vibration modes, $\boldsymbol{u}(\boldsymbol{r}) = \nabla \Psi(\boldsymbol{r})$, in a bottle structure is obtained by solving the wave equation for the scalar potential $\Psi(\boldsymbol{r})$,
$$(\nabla^2 + k_0^2)\Psi(\boldsymbol{r}) = 0, \qquad (1)$$
where $k_0 = \Omega/v_0$ with the vibration frequency, $\Omega$, and the speed of sound wave in silica glass, $v_0$. Solving this differential equation in bottle coordinates $(r, \phi, z)$ with a coordinate vector of $\boldsymbol{r} = (rf(z)\cos\phi, rf(z)\sin\phi, z)$ results in
$$\Psi(r,z) = C_0 J_0(k_r r) H_n(\sqrt{k_0\beta}z) \exp\left[-\frac{k_0\beta}{2}z^2\right]. \qquad (2)$$
where $C_0$ is an arbitrary constant, $\beta$ is microbottle curvature with $f(z) = \sqrt{1-\beta^2 z^2}$, $k_r \equiv \sqrt{k_0^2 - (2n+1)k_0\beta}$ with the axial mode number $n$, and $J_i(\cdot)$ and $H_i(\cdot)$ are the $i$th order Bessel function and Hermite polynomial (see the details in Supplementary Material). Note that this expression is valid under assumptions of no azimuthal distribution and small curvature ($\beta z \ll 1$). Thus, the components of the breathing mode can be readily derived as
$$u_{z,n}(\boldsymbol{r}) = C_0 J_0(k_r r)\left[2n\sqrt{k_0\beta}H_{n-1}(\sqrt{k_0\beta}z) - k_0 z\beta H_n(\sqrt{k_0\beta}z)\right]\exp\left[-\frac{k_0\beta}{2}z^2\right], \qquad (3)$$
$$u_{r,n}(\boldsymbol{r}) = -C_0 k_r J_1(k_r r) H_n(\sqrt{k_0\beta}z) \exp\left[-\frac{k_0\beta}{2}z^2\right].$$
The mechanical frequency $\Omega$ is determined by taking into account the free boundary condition $\partial_r u_r(r,z)|_{r=R_0} = 0$ on the microbottle surface $r = R_0$ with the maximal microbottle radius, $R_0$. This approximated boundary condition corresponds to $J_0(s_i) - J_2(s_i) = 0$, where $s_i$ are the $i$-

th root of this equation. The mechanical frequency $\Omega_{j,n}$ of the $j$th radial breathing and $n$th axial breathing mode is expressed by

$$\Omega_{j,n} \approx v_0 \sqrt{\left(\frac{s_j}{R_0}\right)^2 - \left(n + \frac{1}{2}\right)^2 \beta^2} + (n + 1/2)\beta v_0. \tag{4}$$

The spatial distributions for different axial mode numbers are shown in the Supplementary Material.

**Estimation of effective mass**
The definition of effective mass is given by

$$m_{\text{eff}} = \rho_0 \int dV \, |\phi(\boldsymbol{r})|^2. \tag{5}$$

where $\rho_0$ is the density of silica glass, and $\phi(\boldsymbol{r})$ is the spatial distribution of mechanical modes normalized as $\max_{\boldsymbol{r}} |\phi(\boldsymbol{r})| = 1$. By using Eq. (3), $m_{\text{eff}} = 2.7 \times 10^{-8}$ kg is achieved for the 1st radial and 2nd axial vibrating modes. From this value, we can also estimate the power spectral density of thermal fluctuation $S_{\text{th}} = \sqrt{2\hbar n_{\text{th}}/m_{\text{eff}}\Omega\Gamma} = 1.1 \times 10^{-17}$ m/$\sqrt{\text{Hz}}$, where $n_{\text{th}} = 2.0 \times 10^5$ is the phonon number at room temperature, and $\Omega$ and $\Gamma$ are the frequency and linewidth of the mechanical modes.

**Coupled mechanical mode theory**
To fit the experimentally observed thermal fluctuation in the coupled mechanical modes, we consider the equation of motion for the displacement in the upper and lower microbottles, $x_U$ and $x_L$, respectively, as follows:

$$\begin{aligned}\ddot{x}_U + \Gamma_U \dot{x}_U + \Omega_U^2 x_U + g_M \Omega_U x_L &= f_{\text{th,U}}, \\ \ddot{x}_L + \Gamma_L \dot{x}_L + \Omega_L^2 x_L + g_M \Omega_L x_U &= f_{\text{th,L}},\end{aligned} \tag{6}$$

where $\Gamma_i$, $\Omega_i$, and $f_{\text{th},i}$ are the mechanical damping, mechanical frequency, and Langevin force of the upper (i=U) and lower (i=L) microbottles, and $g_M$ is a mechanical coupling constant. Note that we assume that the Langevin force exerted on each microbottle is independent, i.e., $\langle f_{\text{th},i}(t) f_{\text{th},j}(s) \rangle = \delta(t-s)\delta_{ij} 2k_B T \Gamma_j/m$ with the mechanical mass $m$, which is approximately equivalent between the two microbottles. Because the displacement in the microbottle in air is optomechanically detected, we theoretically formulate it by solving Eq. (6) in the frequency domain as follows:

$$\sqrt{\langle x_U^2(\omega) \rangle} = \sqrt{\frac{2k_B T}{m} \left[ \frac{\Gamma_L g_M^2 \Omega_U^2}{|\chi_U(\omega)\chi_L(\omega) + g_M^2 \Omega_U \Omega_L|^2} + \frac{\Gamma_U |\chi_L(\omega)|^2}{|\chi_U(\omega)\chi_L(\omega) + g_M^2 \Omega_U \Omega_L|^2} \right]}, \tag{7}$$

where $\chi_j(\omega) = (-\omega^2 + \Omega_j^2) + i\Gamma_j \omega$ is the mechanical susceptibility. By fitting this function to the experimental results in Fig. 2D, we can estimate the coupling constant $g_M$, the initial detuning $\delta_M = (\Omega_U - \Omega_L)/2\pi$, and the cooperativity $C_M = \frac{g_M^2}{\Gamma_U \Gamma_L}$, which determines the performance of signal transduction from the lower microbottle to the upper one.

**Fluid-structure interaction in microbottle structures**
The fluid-structure interaction is investigated by referring to that in microdisk resonator (7). Here, we simply assume that the target medium is an incompressible viscous one. The mechanical frequency shift and linewidth broadening due to viscous liquids are given by the following formula:

$$\delta\Omega = \Omega \frac{\iint_{\text{surface}} dS \, [B_r u_r^2 + B_z u_z^2]}{\rho \iiint_{\text{total}} dV \, [u_r^2 + u_z^2]}. \tag{8}$$

$$\delta\Gamma = \Omega \frac{\iint_{\text{surface}} dS\ [A_r u_r^2 + A_z u_z^2]}{\rho \iiint_{\text{total}} dV\ [u_r^2 + u_z^2]}, \tag{9}$$

where $u_r = u_{r,n}(r,z)$ and $u_z = u_{z,n}(r,z)$ are the spatial distributions of microbottles given in Eq. (3). Note that we assume that the radial (axial) displacement only contributes to the longitudinal (shear) interaction to the medium. Thus, each coefficient, $A_j, B_j$ (j = r, z), is approximately given by

$$\begin{aligned} A_r = B_r &\approx \frac{5}{4}\sqrt{2\rho\mu\Omega}, \\ A_z = B_z &\approx \frac{1}{2}\sqrt{2\rho\mu\Omega}. \end{aligned} \tag{10}$$

with a condition $\sqrt{\frac{\mu}{2\rho\Omega}}\frac{1}{R_0} \ll 1$, where $\rho$ and $\mu$ are the density and viscosity of the medium, $\Omega$ is the vibration frequency, and $R_0$ is the maximal microbottle radius. Here, we note that the formalism of the mechanical impedance in a spherical structure (i.e., the sphere diameter is regarded as the maximum microbottle diameter, $R_0$) is utilized to determine the coefficient of 5/4 and 1/2 in Eq. (10) in the same manner as for a microdisk system (7). The amount of frequency shift and linewidth broadening is evaluated by taking into account the immersion depth $z_{\text{im}}$ as follows:

$$\delta\Omega = \delta\Gamma \propto \int_{-\infty}^{z_{\text{im}}} dz [R_0(1-\beta^2 z^2)(B_r^L u_r^2 + B_z^L u_z^2)] \\ + \int_{z_{\text{im}}}^{\infty} dz [R_0(1-\beta^2 z^2)(B_r^A u_r^2 + B_z^A u_z^2)], \tag{12}$$

where $B_k^j$ is the coefficient with the liquid properties (k = L) and the air properties (k = A), and the factor $R_0(1-\beta^2 z^2)$ comes from the Jacobian of the bottle coordinates. Utilizing the expression of spatial distribution of $u_r$ and $u_z$, we can theoretically fit the mechanical frequency and linewidth broadening in Fig. 3D and 3E, respectively.

**Experimental setup for measuring optomechanically induced amplification**
The basic setup was the same as that used for tracking thermal fluctuations (see the schematic in Fig. 4A). Instead of using the electric spectrum analyzer to measure the output radio-frequency signal from the avalanche photodiode (APD), it was measured by a vector network analyzer (VNA) with the radio-frequency drive connected to the electro-optic modulator (EOM). The VNA quantifies the real and imaginary part of the detected $S_{21}$ signal, $s_R$ and $s_I$, respectively. The amplitude and phase are calculated from these signals as $\sqrt{s_R^2 + s_I^2}$ and $\tan^{-1} s'_I/s'_R$, where $s'_j$ (j = R, I) is the post-processed signals by subtracting background tilting due to the optomechanical interactions.

**References**
1. Aspelmeyer, M., Kippenberg, T. J., & Marquardt, F. (2014). Cavity optomechanics. Reviews of Modern Physics, 86(4), 1391.
2. Anetsberger, G., Arcizet, O., Unterreithmeier, Q. P., Rivière, R., Schliesser, A., Weig, E. M., Kotthaus, J. P., & Kippenberg, T. J. (2009). Near-field cavity optomechanics with nanomechanical oscillators. Nature Physics, 5(12), 909-914.
3. Hill, J. T., Safavi-Naeini, A. H., Chan, J., & Painter, O. (2012). Coherent optical wavelength conversion via cavity optomechanics. Nature communications, 3(1), 1-7.
4. Park, Y. S., & Wang, H. (2009). Resolved-sideband and cryogenic cooling of an optomechanical resonator. Nature physics, 5(7), 489-493.

**Acknowledgments**

This work was partly supported by JSPS KAKENHI (21H01023).

**Figures and Tables**

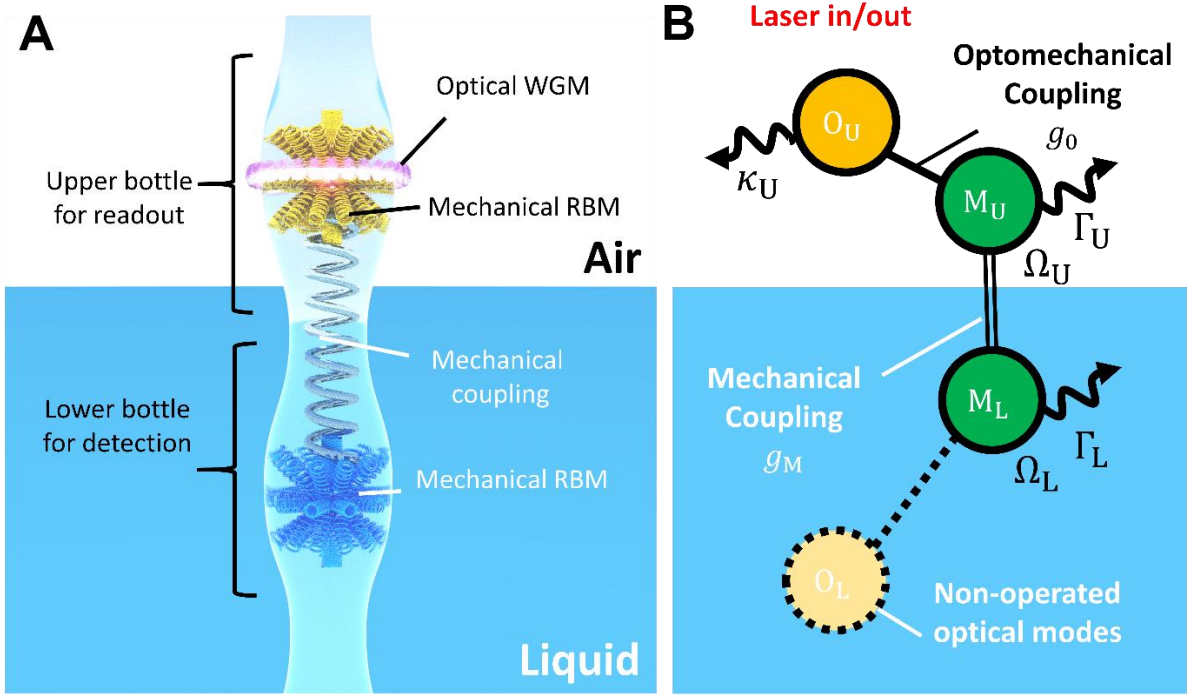

**FIG. 1.**
**Mechanical and optomechanical coupling in TMBR.** (A) Conceptual illustration of the TMBR partially immersed in liquid. An optical whispering gallery mode (WGM) couples to a mechanical radial breathing mode (RBM) in the upper microbottle in air. The mechanical RBMs in the upper and lower microbottles are mutually coupled via the neck part. (B) Schematic illustration of the corresponding optomechanical system. The mechanical mode $M_U$ in the upper microbottle couples to the mechanical mode $M_L$ in the lower microbottle as well as to the optical mode $O_U$ in the upper microbottle. The optical mode in the lower microbottle ($O_L$) is not excited here. The mechanical mode frequency and intrinsic damping rate of each resonator are denoted by $\Omega_i$ and $\Gamma_i$ ($i = U, L$), respectively, and the optical decay rate of the upper microbottle is denoted by $\kappa_U$. The vacuum optomechanical coupling rate and the mechanical coupling rate are denoted by $g_0$ and $g_M$, respectively.

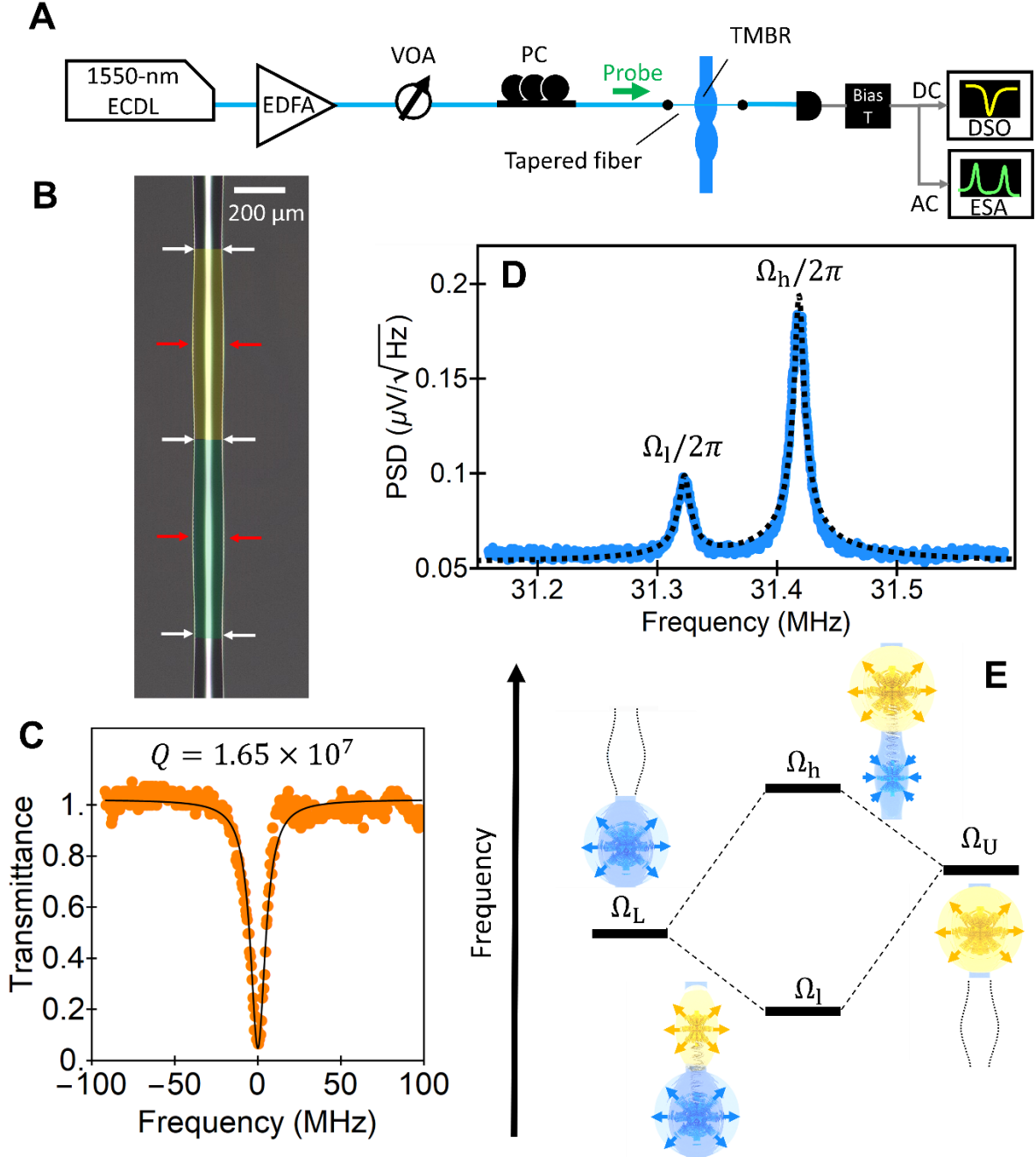

**FIG. 2.**
**Fundamental properties of the TMBR in air.**
(A) Schematic image of the measurement setup. (B) Optical micrograph of the TMBR. The yellow and green shaded area correspond to the two microbottle structures. The white and red arrows respectively indicate the position of the necks and the center of each microbottle. (C) Optical transmission spectrum of the upper microbottle measured with the probe power of 1.8 mW. (D) Thermal noise spectrum of the coupled mechanical modes measured via the frequency noise of the optical transmission at the slope of the cavity resonance shown in C. The broken curve corresponds to the theoretical fit with the double Lorentzian obtained by the coupled mode theory. (E) Schematic energy diagram that depicts the coupling of the energy levels of the upper and lower microbottles.

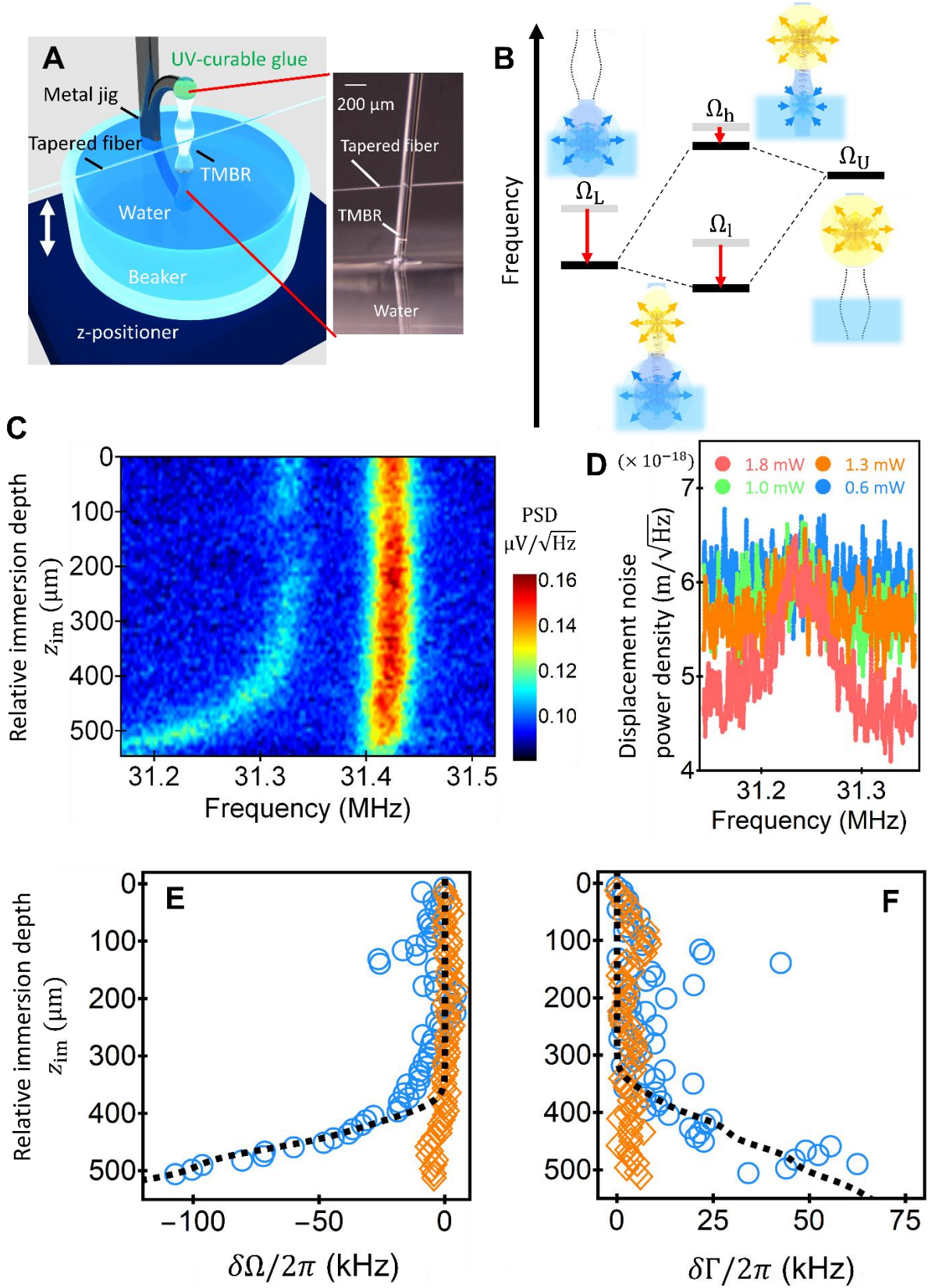

**FIG. 3.**
**Optical probe of thermal motion of the microbottle in water.** (A) Photographic image of the setup for the dip-in-water measurements. (B) Schematic energy diagram that depicts the shift of the coupled mechanical modes in water. (C) Immersion depth dependence of the thermal noise spectrum of the coupled mechanical modes. The vertical axis corresponds to the relative immersion depth, $z_{im}$, which is defined as the depth from the initial position of the liquid surface close to the bottom neck (indicated by the bottom white arrows in Fig. 2B). (D) Thermal noise spectrum with respect to the probe optical power. (E) and (F) Frequency shift and linewidth change with respect to $z_{im}$, respectively. The broken curves are the theoretical fitting given by taking into account of the fluid-structure interaction in an incompressible viscous medium.

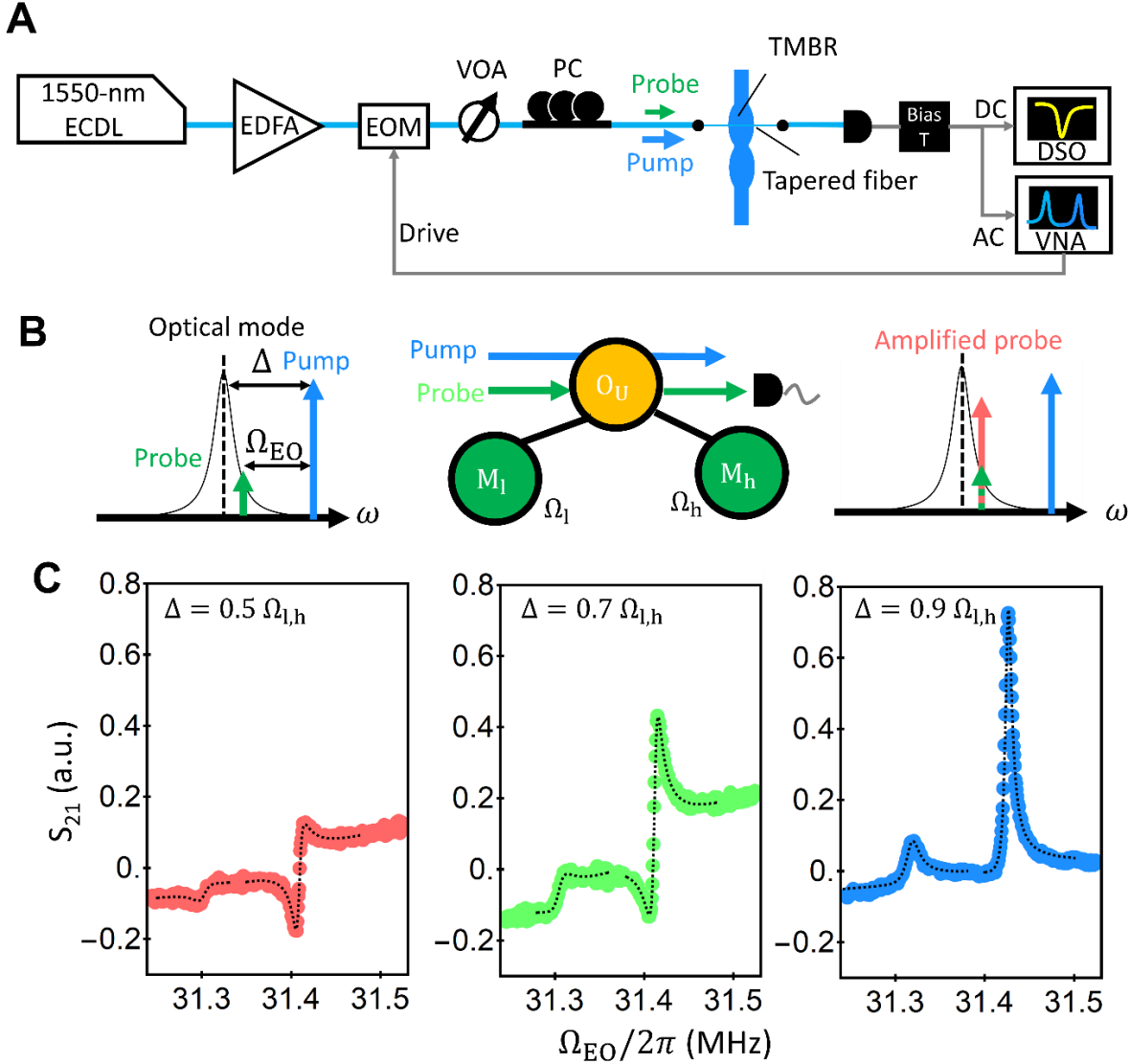

**FIG. 4.**
**Cavity optomechanical operation of the microbottle in the resolved sideband regime.** (A) Schematic image of the measurement setup. (B) Conceptual illustration of the blue-detuned pumping, which leads to the optomechanically induced amplification. (C) Frequency response of the coupled two mechanical modes measured through the optical transmission ($S_{21}$) of the probe light for three different blue-detunings $\Delta = 0.5$, 0.7 and 0.9 $\Omega_{h,l}$ at $z_{im} = 0$ μm, where $\Delta$ was extracted from the theoretical fitting (broken curves).

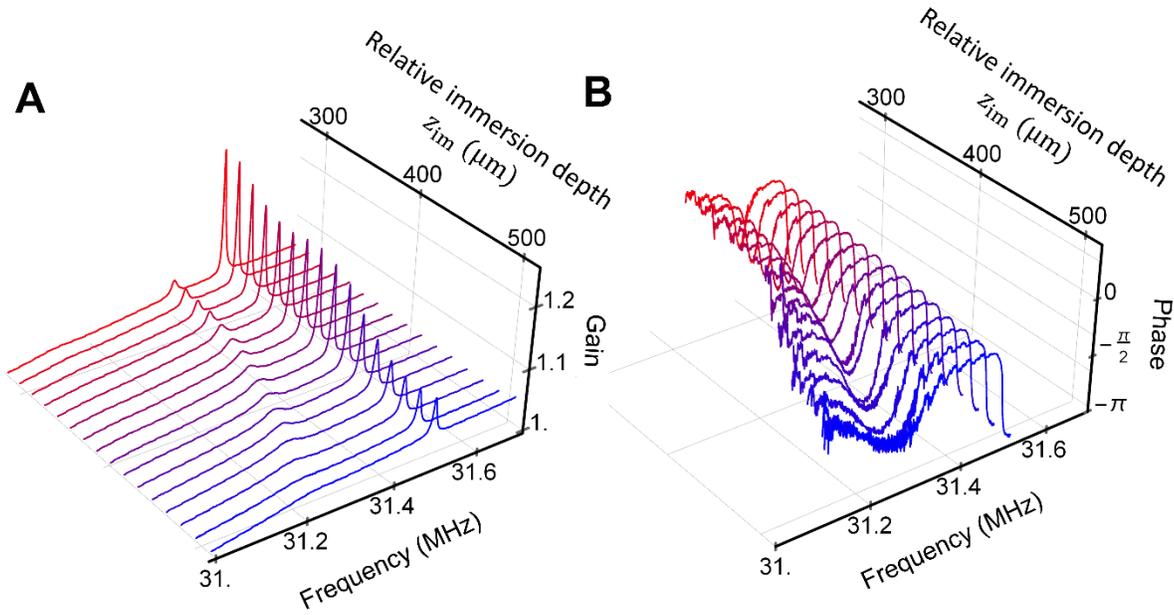

**FIG. 5.**
**Optomechanically induced amplification partially immersed in the water.**
(A) and (B) Optomechanically induced amplification by immersing the lower bottle into the water. The dependence of $z_{im}$ on the amplitude (A) and the phase (B) of the coupled mechanical modes under the blue-detuned pumping of $\Delta = 0.9\Omega_{h,l}$.

### The optical quality factor ($Q$) under the dip-in-water condition
Figure S1 shows the immersion depth dependence of the optical $Q$. It is confirmed that the optical $Q$ is maintained even under the dip-in-water condition. This is because the optical mode formed in the upper microbottle (in air) is optically isolated from the lower microbottle (in water).

### Several mechanical modes appeared near 31 MHz
Figure S2 shows the wide range of the thermal noise spectrum near 31 MHz, measured via the frequency noise of the optical transmission at the slope of the cavity resonance. Several pairs of resonance peaks correspond to the coupled mechanical radial breathing modes of the two microbottles with no node along the radial direction but with different node numbers with respect to the axial direction.

### Analytical expression of vibration mechanical mode in microbottle structure
To solve the differential equation for the scalar potential for the vibration mode,
$$(\nabla^2 + k_0^2)\Psi(\mathbf{r}) = 0, \qquad (S1)$$
under a microbottle coordinate $\mathbf{r} = (rf(z)\cos\phi, rf(z)\sin\phi, z)$ with $f(z) = \sqrt{1-\beta z^2}$, we explicitly expand the Laplacian in the microbottle coordinate as
$$\nabla^2 = \frac{1}{f^2(z)}\left[\left(1 + r^2(f'(z))^2\right)\partial_r^2 + \frac{1 - r^2 f(z)f''(z)}{r}\partial_r - 2rf(z)f'(z)\partial_r\partial_z + \frac{1}{r^2}\partial_\phi^2 \right. \qquad (S2)$$
$$\left. + f(z)^2 \partial_z^2\right].$$
Assuming the adiabatic curvature condition, i.e., $\beta z \ll 1$, Eq. (2) can be reduced to
$$\nabla^2 \approx \frac{1}{f^2(z)}\left[\partial_r^2 + \frac{1}{r}\partial_r + \frac{1}{r^2}\partial_\theta^2 + f^2(z)\partial_z^2\right]. \qquad (S3)$$

By taking into account the radial breathing mode $\Psi(\mathbf{r}) = \Psi(r,z)$ which is azimuthally symmetric, the differential equation simply becomes

$$\left[\partial_r^2 + \frac{1}{r}\partial_r + f^2(z)\partial_z^2 + k_0^2\right]\Psi(r,z) = 0. \tag{S4}$$

From the separation $\Psi(r,z) = R(r)Z(z)$,

$$\partial_z^2 Z(z) + [k_0^2 - k_r^2 - k_0^2\beta^2 z^2]Z(z) = 0, \tag{S5}$$
$$[r^2\partial_r^2 + r\partial_r + k_r^2 r^2]R(r) = 0. \tag{S6}$$

Thus, we can represent $R(r)$ and $Z(z)$ by the spatial functions

$$Z(z) \propto H_n\left(\sqrt{k_0\beta}z\right)\exp[-k_0\beta z^2/2], \tag{S7}$$
$$R(r) \propto J_0(k_r r), \tag{S8}$$

where $H_n(\cdot)$ and $J_0(k_r r)$ are the $n$th order Hermite polynomial and the zeroth order Bessel function with $k_r = \sqrt{k_0^2 - (2n+1)k_0\beta}$. The scalar potential of vibration mode is given by

$$\Psi(r,z) = C_0 J_0(k_r r) H_n\left(\sqrt{k_0\beta}z\right)\exp[-k_0\beta z^2/2] \tag{S9}$$

with an arbitrary constant $C_0$.

The vector representation of displacement fields $\mathbf{u}(r,z) = \nabla\Psi$ is calculated as

$$u_{z,n}(\mathbf{r}) = C_0 J_0(k_r r)\left[2n\sqrt{k_0\beta}H_{n-1}\left(\sqrt{k_0\beta}z\right) - k_0 z\beta H_n\left(\sqrt{k_0\beta}z\right)\right]\exp\left[-\frac{k_0\beta}{2}z^2\right], \tag{S10}$$

$$u_{r,n}(\mathbf{r}) = -C_0 k_r J_1(k_r r) H_n\left(\sqrt{k_0\beta}z\right)\exp\left[-\frac{k_0\beta}{2}z^2\right].$$

with an approximation $\nabla = \partial_r \mathbf{e_r} + \partial_z \mathbf{e_z}$

## Spatial distributions of mechanical modes with different axial mode numbers

The spatial distributions of mechanical displacement $u_{r,n}(r,z)$ and $u_{z,n}(r,z)$ and the vibrating energy $\mathcal{E}(r,z) = u_{r,n}^2 + u_{z,n}^2$ from Eq. (S10) are represented in Fig. S3 with the axial mode number $n = 0, 1,$ and 2.

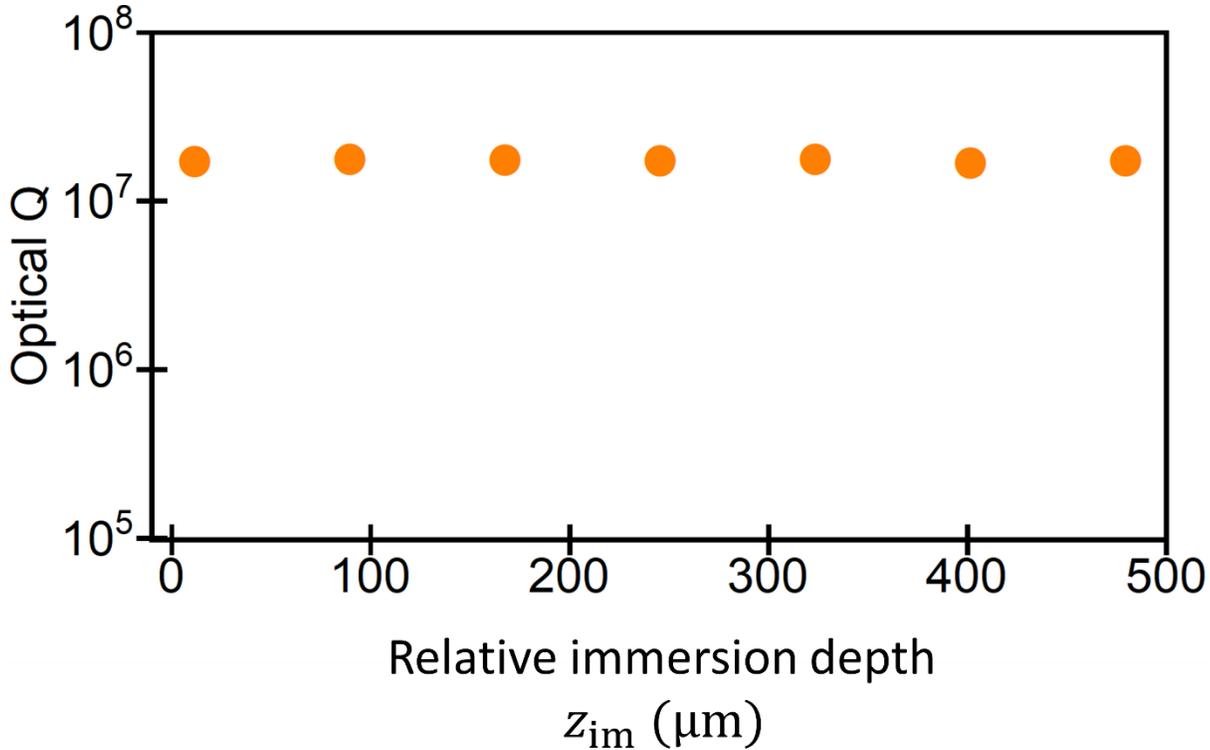

**Fig. S1** Optical $Q$ with respect to the immersion depth to water.

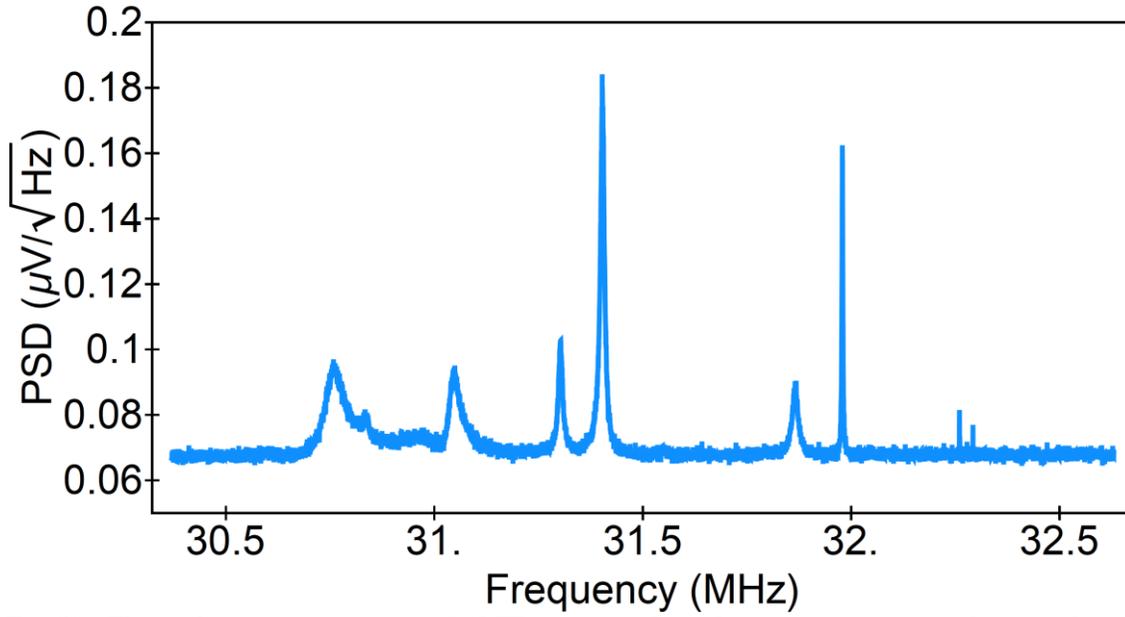

**Fig. S2** Thermal noise spectrum near 31 MHz, measured via the frequency noise of the optical transmission at the slope of the cavity resonance.

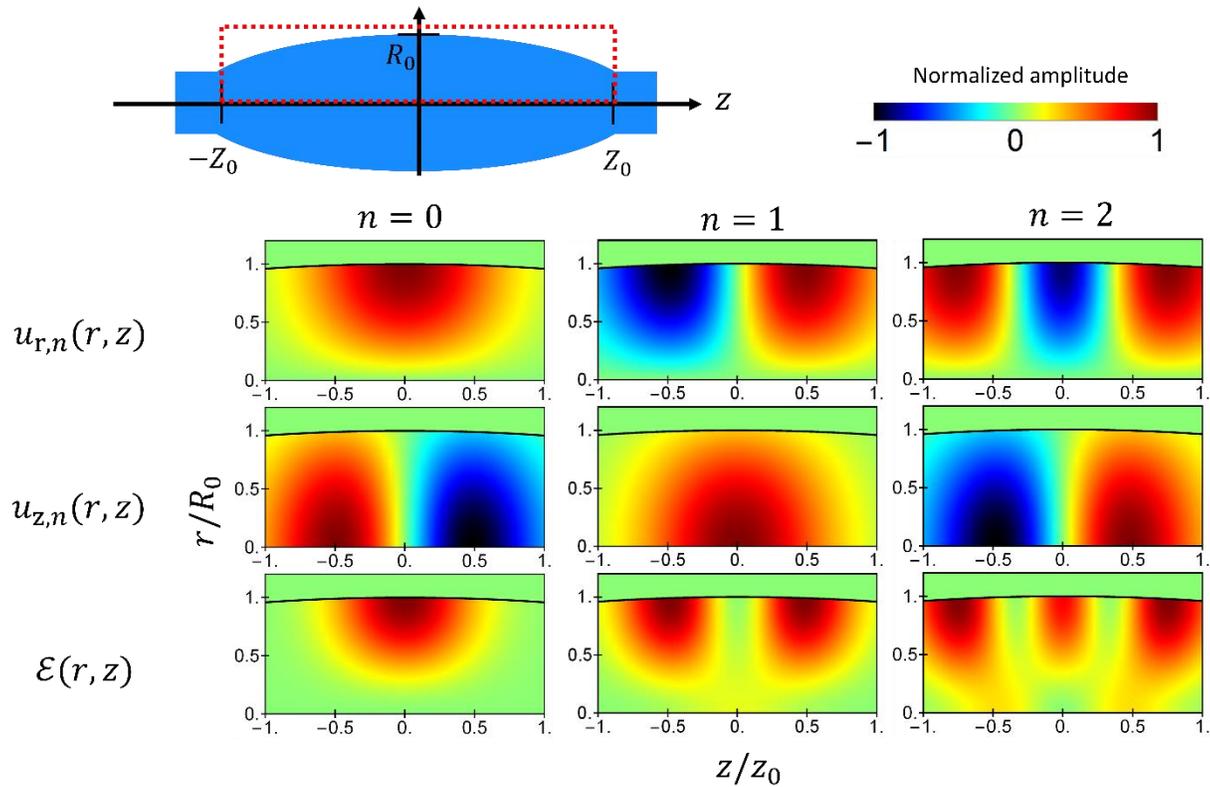

**Fig. S3 Spatial distributions of mechanical displacement and energy in different axial mode numbers.** The red dotted rectangle in the upper schematic shows the plotted region, i.e., the main part of the microbottle structure. The vertical and horizontal axes in the density plots are scaled by the maximum dimeter, $R_0$, and the z coordinate of the necks, $Z_0$. The solid black curve in the density plot shows the bottle surface.